\documentclass[aps,pre,preprint,amssymb,amsmath,nobibnotes,showpacs,floatfix]{revtex4}
\usepackage{graphicx}
\usepackage{dcolumn}
\usepackage{bm}
\begin{document}
\title{Long time behavior of quasi-stationary states of the Hamiltonian Mean-Field model}
\author{Alessandro Campa}
\affiliation{Complex Systems and Theoretical Physics Unit, Health and Technology Department,
Istituto Superiore di Sanit\`a, \\ and INFN Roma1, Gruppo Collegato Sanit\`a,
Viale Regina Elena 299, 00161 Roma, Italy}
\author{Andrea Giansanti}
\author{Gianluca Morelli}
\affiliation{Physics Department, Universit\`a di Roma ``La Sapienza'' \\
Piazzale Aldo Moro 2, 00185 Roma, Italy}
\date{\today}
\begin{abstract}
The Hamiltonian Mean-Field model has been investigated, since its introduction about a decade ago,
to study the equilibrium and dynamical properties of long-range interacting systems. Here we study
the long-time behavior of long-lived, out-of-equilibrium, quasi-stationary dynamical states, whose
lifetime diverges in the thermodynamic limit. The nature of these states has been the object of a
lively debate, in the recent past. We introduce a new numerical tool, based on the fluctuations of the
phase of the instantaneous magnetization of the system. Using this tool, we study the quasi-stationary
states that arise when the system is started from different classes of initial conditions, showing that
the new observable can be exploited to compute the lifetime of these states. We also show that
quasi-stationary states are present not only below, but also above the critical temperature of the
second order magnetic phase transition of the model. We find that at supercritical temperatures the
lifetime is much larger than at subcritical temperatures.
\end{abstract}
\maketitle

\section{Introduction}

Many examples of long-range interacting systems can be found: self-gravitating systems \cite{padma},
unscreeened Coulomb systems \cite{bryd}, trapped charged particles \cite{dubin}, wave-particle
interactions \cite{elsk}, vortices in two-dimensional fluid mechanics \cite{cha1}, magnets where
dipolar effects are dominant \cite{engl}. The study of both equilibrium and out-of-equilibrium
properties of systems with long-range interactions poses several challenges, that in recent years
have been faced through analytical and numerical methods (for a review see, e.g., Ref. \cite{draw}).
It has been shown that different statistical ensembles can be nonequivalent, so that the equilibrium
states which can be reached by fixing certain thermodynamic parameters may be different from those
obtained fixing other thermodynamic parameters. Rigorous results have been produced in this
field \cite{elhatu}. The approach to equilibrium reveals the existence of
transient states whose lifetime can diverge in the thermodynamic limit (i.e., when the number $N$ of
degrees of freedom of the system goes to infinity): these states can be called quasi-stationary states
(QSS), and it is worth underlining that they are not metastable states, i.e., they are not stable local
extrema of thermodynamic potentials, but their robustness is of dynamical origin \cite{draw}. Moreover,
the characteristics of the QSS can depend on the initial conditions of the system. Another very interesting dynamical property is the breaking of ergodicity in microcanonical dynamics \cite{borgo,mukam}. These
facts imply that a deep understanding, in long-range systems, of kinetic effects, and in particular of
the features of QSS, can be achieved considering both thermodynamics and dynamics and their intricate
relationship.

This program has been pursued for several years on a simple model originally introduced in
Ref. \cite{antruf} and called the Hamiltonian Mean-Field (HMF) model. The model is an approximation
of one-dimensional gravitational interactions, and it is also closely related to the Colson-Bonifacio
model for free electron laser \cite{baretal}. The Hamiltonian is:
\begin{equation}\label{ham}
H = K +  V = \frac{1}{2}\sum_{i=1}^N p_i^2 + 
\frac{1}{2N} \sum_{i,j=1}^{N} \left[1 - \cos(\theta_i - \theta_j )\right]\, .
\end{equation}
It refers to a system of $N$ globally coupled rotators of unit mass, each one being described by the
angle variable $\theta_i$ ($-\pi \leq \theta \leq \pi$) and by its conjugate momentum $p_i$ (that in
the following, for short, will be denoted as the velocity). The coupling constant
is scaled by the number of rotators. This quite unphysical rescaling (the Kac prescription \cite{kac})
is necessary to ensure extensivity of the thermodynamic potentials, but it is not dramatic in the study
of dynamical properties, since, as long as $N$ is not infinite, it is equivalent to a rescaling of time.

The statistical mechanics of this system can be exactly solved, both in the canonical \cite{antruf} and
in the microcanonical \cite{bbdr} ensembles, that, for this model, have been shown to be equivalent. The
system has a ferromagnetic second order phase transition at a critical temperature $T_c=0.5$, corresponding
to a critical energy per particle (or energy density) $E_c/N=\epsilon_c=0.75$. The magnetization, that
spontaneously attains a nonzero value below the critical temperature, is the modulus $M$ of the ensemble
average of the vector:
\begin{equation}
\label{magdef}
\mathbf{m} = (m_x,m_y) \equiv \frac{1}{N}(\sum_{i=1}^N \cos\theta_i, \sum_{i=1}^N \sin\theta_i) \, .
\end{equation}
i.e., $M=|\mathbf{M}|$, with $\mathbf{M}=\langle \mathbf{m}\rangle$, is positive below the critical
temperature. The lower bound for the energy density is $\epsilon=0$.

Contrary to the equilibrium case, the out-of-equilibrium behavior of the system presents a great richness.
This work aims at presenting some results concerning the dynamics of the HMF model. We describe the
properties of the QSS that the system exhibits when the initial conditions belong to
different classes. In the remaining of this section we give a short summary of the results, connected
with those presented in this work, that have already been obtained.

Microcanonical molecular dynamics simulations have shown that, for energy densities slightly below the
critical value, QSS are present, whose lifetime diverges with a power of $N$. This implies that, if the
thermodynamic limit $N \rightarrow \infty$ is performed before the infinite time limit, the system remains
trapped in the QSS. While in the QSS, the distribution of the velocities $p_i$ of the rotators is not
Maxwellian \cite{lat01}. The energy density that has been mostly considered is $\epsilon = 0.69$; at
equilibrium $M\approx 0.31$, corresponding to a temperature $T\approx 0.475$. In the simulations the
behavior of $M$ is studied through the observation of the dynamical variable $m=|\mathbf{m}|$, while,
as usual, the temperature is studied through the observation of the dynamical variable $2K/N$, i.e.,
twice the kinetic energy per particle (let us use, for this dynamical variable, the same symbol $T$ of
the thermodynamic temperature). It should be noted that the study of $m$ in the microcanonical simulations
is equivalent to that of $T$, since the conserved Hamiltonian (\ref{ham}) can be written as:
\begin{equation}\label{ham1}
H = \frac{N}{2}T + \frac{N}{2} \left(1 - m^2 \right)\, .
\end{equation}
The simulations have shown that the details of the dynamics in the QSS depend on the initial
conditions. The most studied classes of initial conditions are those in which the initial value of
$m$ is either $0$ or $1$ (obtained with a uniform initial distribution of the $\theta_i$ or setting
all $\theta_i$ equal to zero, respectively), while the velocities $p_i$ are uniformly distributed
in a range whose extension is determined by $\epsilon$. When $m(0)=0$ it has been found that the
lifetime of the QSS diverges as $N^{1.7}$ \cite{yama}, while in the case $m(0)=1$ this divergence is linear
in $N$ \cite{lat01}. In both cases the magnetization $m$ in the QSS converges to zero for increasing $N$,
although differences in the details of this convergence are observed. For example, starting
with $m(0)=1$ the magnetization in the QSS, for a given $N$ value, depends on the initial conditions
(i.e., on the different realizations, for finite $N$, of the uniform velocities distribution, see Fig. 1);
it is then necessary to perform several runs to obtain an average value. Starting from $m(0)=0$ the
different runs are much more similar.

The observed non-Gaussian character of the velocity distributions has given rise to a
lively debate on the characterization of these distributions. In particular, numerical investigations have
concerned the tails of the distributions, to see if their decay is exponential (or even faster) \cite{yama},
or if the decay could be fitted \cite{lat02} to the expressions derived in the framework of nonextensive
thermodynamics \cite{tsal}, that predicts tails decaying with a power law. The controversy has extended
to the study of the anomalous diffusion \cite{bouch,pluc}, and fits to nonextensive thermodynamics
expressions have been done also for the cases where the initial magnetization takes values between $0$
and $1$ and for the model where the coupling between rotators has a slow decay with distance \cite{rapnew}
(see Ref. \cite{campa} for the generalization of the HMF model and a detailed study of its equilibrium
behavior). In this paper we are not directly concerned with the characterization of the QSS in terms of
ordinary Boltzmann-Gibbs (BG) or non extensive thermodynamics, and we limit ourselves to the following
remarks.

Recent analytical calculations have shown that it is possible to interpret the QSS of the HMF
model within a dynamical approach based on the Vlasov equation \cite{yama}. In fact, it has
been proved \cite{braun} that, in the limit $N\rightarrow \infty$ the microscopic one-particle distribution
function obeys this equation for a class of mean-field models, to which HMF belongs. Following the same
line of research, it has then been argued that the QSS are formed in a short time through a fast relaxation
to a state that maximizes an entropy functional of fermionic type \cite{anton,cha2}, similarly to what
happens for gravitational systems \cite{lynd}. The approach to the QSS and its short time behavior seems
well reproduced by this theory, although some details need further explanations \cite{anton}. The long
time behavior, with the slow approach to the final BG equilibrium state, is much less understood from an
analytical point of view, although again the Vlasov equation can be of help in justifying the very slow
relaxation \cite{yama,bouch}.

It is this long time behavior that we are concerned with in this paper. We do not offer new analytical
tools; rather, we study in details the characteristics of the velocity distribution functions of the QSS
that arise when the dynamics starts from several different classes of initial conditions. The main point
is the introduction of a new tool that characterizes the QSS, and that is based on the fluctuations of the
phase of the magnetization. Mainly on the basis of this tool, we show that QSS are present also above the
critical temperature, a fact that, up to now, has been overlooked in the literature.

In Section II we explain, referring also to appendix A, the different classes of initial conditions. In
the following three sections we present our results, focussing, respectively, on the role of the initial
conditions on the properties of the QSS, on the dynamics of the magnetization using the new tool related
to its phase, and on the QSS that occur at supercritical energy densities. The discussion, relating ours
with previously published results, follows in the last section.

\section{Numerical simulations: the different classes of initial conditions}

The equations of motion derived from Hamiltonian (\ref{ham}) can be written as:
\begin{equation}\label{equ}
\ddot{\theta}_i = -m \sin \left(\theta_i -\phi \right) \, ,
\end{equation}
where $m=|\mathbf{m}|$ and $\phi = \arctan (m_y/m_x)$ are the polar coordinates of the vector $\mathbf{m}$.
The equations, that in this form clearly emphasize the mean field character of the system, have been
numerically integrated, in microcanonical simulations, with a fourth order symplectic algorithm \cite{yoshi},
with an integration timestep $dt=0.1$, which ensures an energy conservation with a relative error of the
order of $10^{-5}$.

The initial conditions that have been explored in this work can be characterized by the
one-particle distribution functions $f(\theta,p)$ that the initial values of $\theta_i$ and $p_i$
are intended to realize. In this work we consider initial conditions in which the value $m$ of the
magnetization is initially $0$ or $1$, and a single case where it is $0.3$. In all cases
$f(\theta,p)$ is factorizable as $g(\theta)h(p)$. We refer to appendix A for the expressions of
the different distributions; here we limit ourselves to a few details. As $g(\theta)$
we always consider a distribution function that is constant inside a range and $0$ outside; the range is
determined by the value of $m$ that one wants to set. For $h(p)$ we consider distributions with compact
support, (see appendix A). Among the initial conditions here considered there are the two types that have
been mostly used in the literature: i) the so called water bag (wb) initial conditions, characterized by
$g(\theta)=\delta(\theta)$ (i.e., $m=1$) and $h(p)=1/(2p_{wb})$ between $-p_{wb}$ and $p_{wb}$, with
$p_{wb}$ determined by the value of the energy, and ii) the uniform (u) initial conditions, characterized
by $g(\theta)=1/(2\pi)$ in the entire $\theta$ range (i.e., $m=0$) and $h(p)=1/(2p_{un})$ between $-p_{un}$
and $p_{un}$, with $p_{un}$ again determined by the value of the energy.

\section{The role of the initial conditions in the occurrence and in the properties of
quasi-stationary states}

As mentioned in the Introduction, the microscopic one-particle distribution function $f(\theta,p,t)$ obeys,
in the limit $N\rightarrow \infty$, the Vlasov equation \cite{braun}; in appendix B we give this equation
for the HMF system. For large values of $N$ it is expected that the one-particle distribution will deviate
from the solution of the Vlasov equation because of finite size effects, but that these effects should be
small. It is immediate to see that a distribution uniform in $\theta$ is a stationary solution
of the equation; therefore, if in addition it is possible to prove its stability, one should expect that
for large $N$ such one-particle distribution will be maintained for a long time, giving rise to a QSS.
Uniformity in $\theta$ is not a necessary condition for stable stationarity with respect to the Vlasov
equation, therefore it is possible to find also nonuniform distributions that produce a QSS.
The two questions related to this fact are the following: {\it i)} if the system is prepared in a generic
initial condition, i.e., a state that is not a stable stationary solution of the Vlasov equation, one would
like to know if it reaches such a state, and thus remains in a QSS before eventually going to BG equilibrium;
{\it ii)} when the system is in a QSS, either by preparation or by reaching it from a generic initial
condition, what are the modalities by which the system reaches BG equilibrium, and how the modalities depend
on the preparation of the system. The first question has been recently studied \cite{cha2,anton} for a
simple particular class of initial conditions, that should relax in a short time to a QSS; answers to the
second questions up to now are only of numerical nature, relying on results of simulations, with the
exceptions of some arguments again based on the Vlasov equation \cite{yama,bouch}. The QSS lasts for a
time proportional to a power of $N$, after which the one-particle distribution relaxes to the BG
equilibrium expression:
\begin{equation}\label{bgf}
f(\theta,p) = A \exp\left\{-\frac{\beta}{2}p^2 + \beta M \cos \left(\theta - \phi \right)\right\} \, ,
\end{equation}
where the values of the inverse temperature $\beta$ and of the spontaneous magnetization $M$ are those
computed in the microcanonical or canonical ensembles, and where the normalization factor $A$ is
proportional to $I_0(\beta M)$, the modified Bessel function of order $0$. When $M \ne 0$, the
magnetization phase $\phi$ is determined by the boundary conditions. It is useful
to stress again that the QSS are not thermodynamical metastable states, and therefore their properties
can not be deduced by the study of thermodynamical potentials. In this paper we provide more extended
numerical results on the relaxation to equilibrium of QSS, especially with the introduction of new tools.
 
If the dynamics of the system starts from random $\theta_i(0)$ and $p_i(0)$, it usually does not get trapped
into a QSS, so that the one-particle distribution function rapidly reaches the form (\ref{bgf}), then
the temperature and average magnetization attain their BG values. Only the preparation in selected non
equilibrium initial conditions induces a dynamics that generates a QSS. This is palusible, especially on
the basis of point {\it i)} treated above: apart the case in which the system is prepared in a stable
stationary solution of the Vlasov equation, it is not expected that the rapid early evolution of a generic
initial evolution will lead to such a solution.

Usually, the QSS have been studied at energy densities $\epsilon$ slightly below the critical value $0.75$,
and mostly at the value $0.69$. However, it seems that there is no argument that prevents QSS from
occurring also above the critical value $0.75$. In fact, in this work we observe that QSS exist even
at supercritical energy densities.

\subsection{Water bag initial conditions}

In the water bag (wb) initial conditions, as put in evidence in the corresponding distributions in
appendix A, the initial angles are all set to zero and the initial velocities are sampled from a
uniform distribution centered on zero; the initial configuration has magnetization $m=1$, and therefore
$T=2\epsilon$, as can be seen from Eq. (\ref{ham1}).
Simulations at energy densities in the range between $0.68$ and $0.75$ have shown that the temperature
falls onto a non equilibrium plateau value $T_{qss}$ within a few timesteps \cite{lat01}. However, the
$N$ dependent value of $T_{qss}$ is an average over many trajectories, i.e., over many realization of the
(wb) initial conditions. For example, if $\epsilon= 0.69$, the canonical solution of the model gives an
equilibrium temperature $T_{eq}=0.475$, while $T_{qss} < T_{eq}$. Correspondingly (see Eq. (\ref{ham1})),
the initial unitary magnetization gets very rapidly a small value, close to zero, that characterizes the
QSS, and, after a time diverging with $N$, reaches the equilibrium value $M = 0.31$. A characteristic of
the (wb) initial conditions is the presence of large sample to sample fluctuations in the relevant
observables (temperature and magnetization). Figs. 1, 2 and 3 illustrate this intrinsic randomness in
the typical case of $\epsilon=0.69$: Fig. 1 shows the temperature time course for $20$ different
trajectories with $N=1000$, together with the average time course; Fig. 2 plots the dispersion, at
different times, of the temperature: the dispersion shrinks only when the system relaxes to BG
equilibrium. Fluctuations around the average temperature of the QSS decrease with increasing $N$; in
addition, when $N$ increases, the average temperature of the QSS decreases and tends to the value $0.38$,
corresponding, according to Eq. (\ref{ham1}), to zero magnetization. Fig. 3 puts in evidence the decrease
of the average QSS temperature when $N$ is increased, and it shows, for the largest $N$ value, the time
fluctuation of the QSS temperature of a single trajectory. In ref. \cite{pluc2} one can find a study of
the dependence of the average QSS temperature as a function of the initial magnetization $m$.
\begin{figure}[htbp!]
\begin{center}
\includegraphics[scale=0.3]{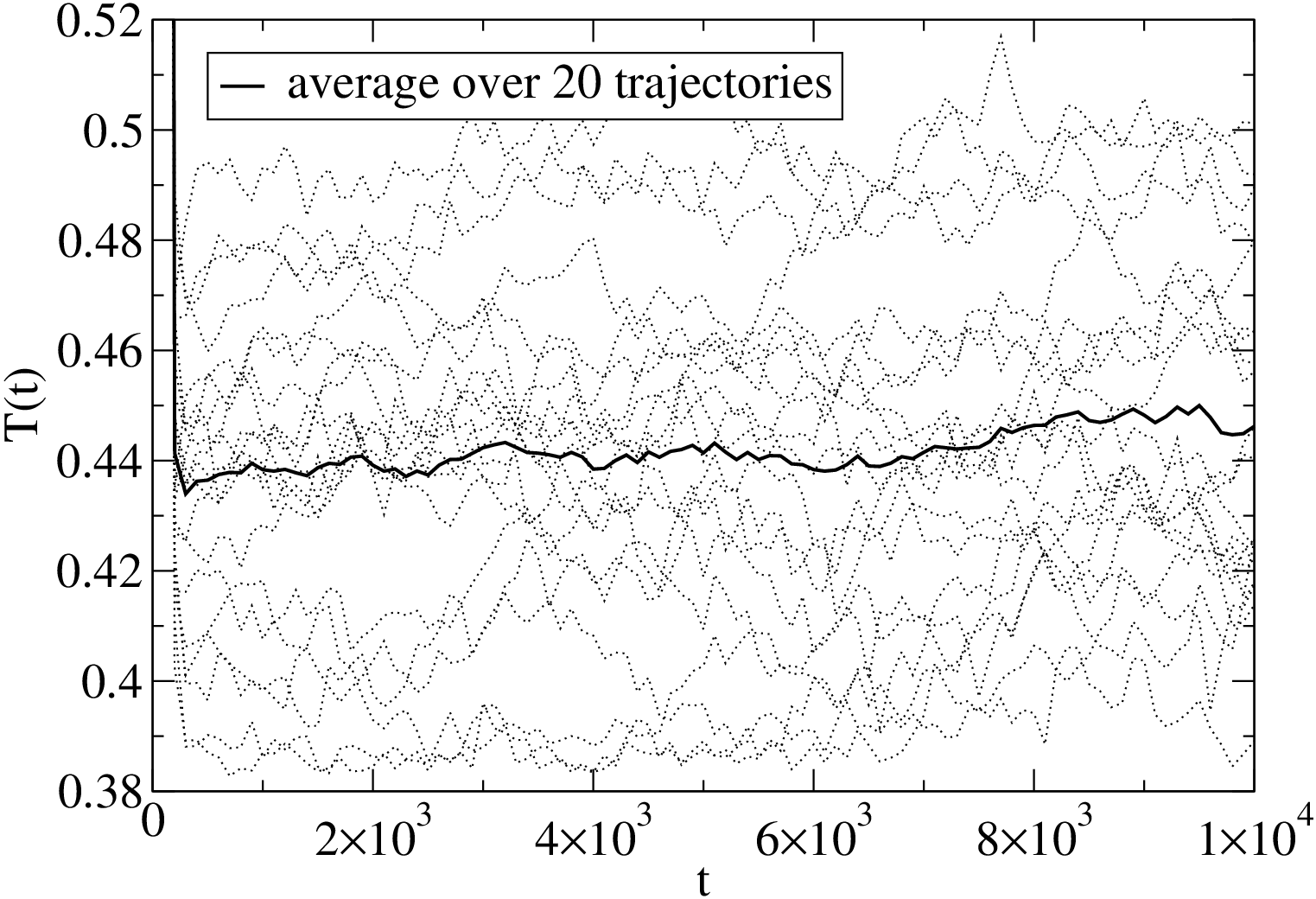}
\caption[]
{Time evolution of temperature in the HMF model with $N=1000$, $\epsilon=0.69$, (wb) initial conditions. The
full bold line is the average over 20 trajectories. The dotted lines refer to the individual trajectories,
that can considerably deviate from the average. Only considering a larger and larger number
of rotators the fluctuations around the average tend to reduce.}
\end{center}
\end{figure}
\begin{figure}[htbp!]
\begin{center}
\includegraphics[scale=0.3]{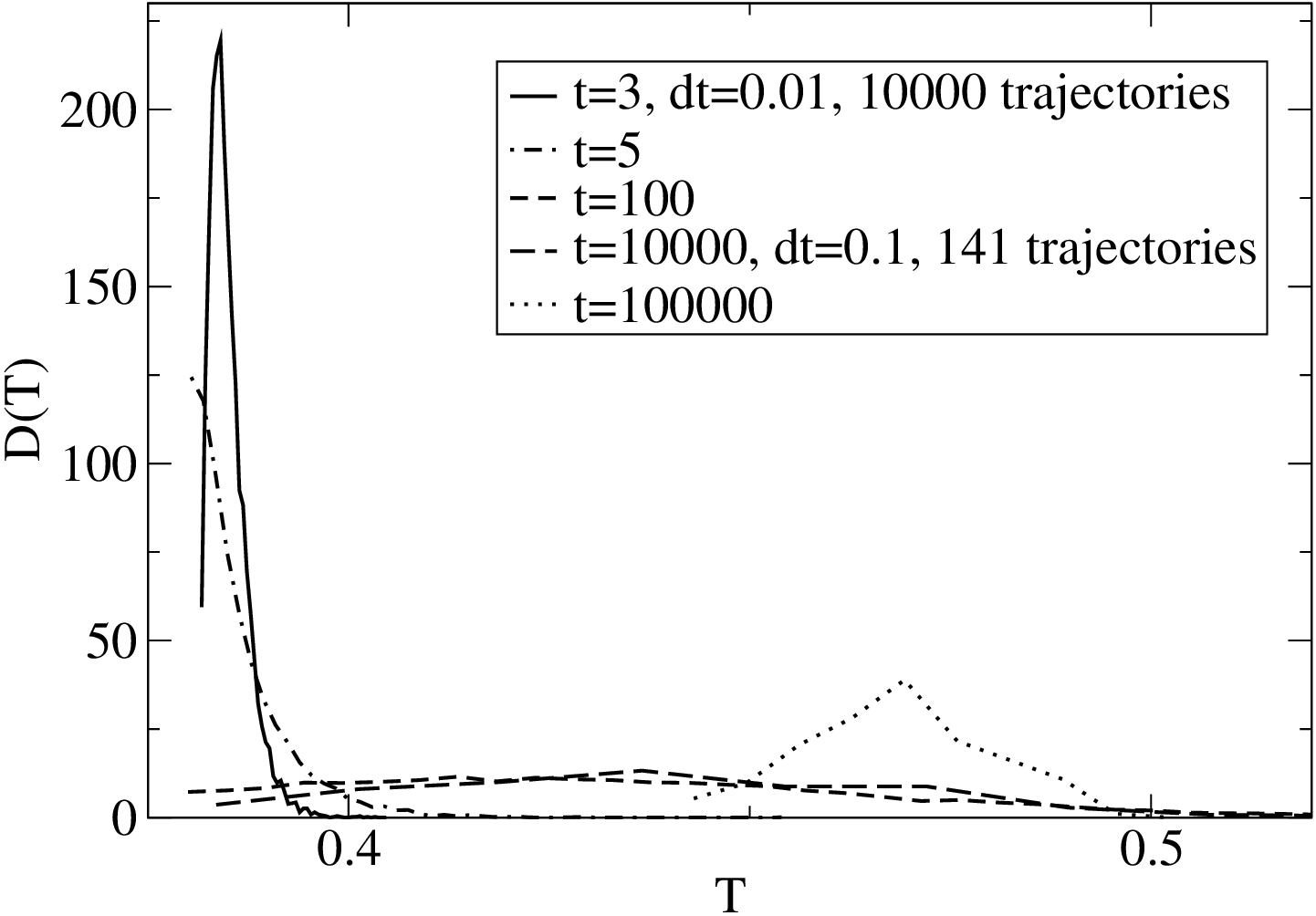}
\caption[]
{Dispersion of the temperature around the average in the HMF model, (wb) initial conditions.
Data refer to the case of $N=1024$ rotators, $\epsilon=0.69$; each dot is an average over 100 timesteps. Note
that, for this number of rotators, after the spread from the initial value (curves corresponding to $t=3$
and $t=5$), temperatures fall in a large interval up to times $10000$ (approximately, the lifetime of the
quasi-stationary state). Only over time $100000$, when the system relaxes toward equilibrium, they tend
to concentrate around the canonical value of $0.475$, expected for the energy density here considered.
Short trajectories have been simulated with a smaller timestep, as indicated.}
\end{center}
\end{figure}
\begin{figure}[htbp!]
\begin{center}
\includegraphics[scale=0.3]{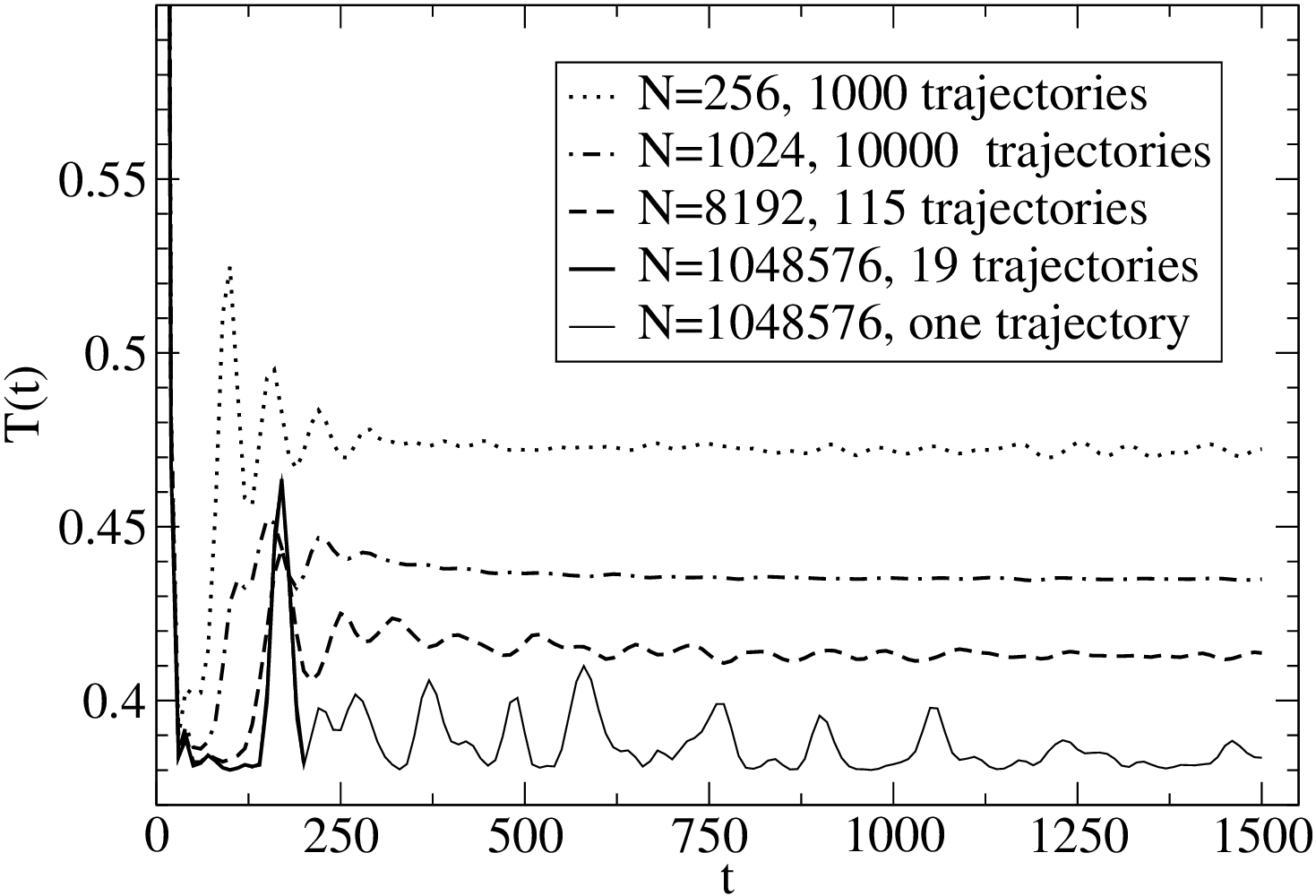}
\caption[]
{Time evolution of the average temperature as a function of $N$, in the HMF model, for the (wb) initial
conditions. Note that increasing the number of rotators the temperature of the quasi-stationary state tends
to the value of $0.38$, corresponding to a zero magnetization state with energy density $\epsilon=0.69$.}
\end{center}
\end{figure}

\subsection{Other classes of initial conditions and the attracting velocity distribution function}

The (wb) initial conditions are implemented with a random extraction of the initial velocities
from a uniform distribution centered around zero. However, it is possible to assign deterministically the
initial velocities with the prescription: $p_i = -p_{wb}+2p_{wb}\left(i-1/2\right)/N$, $i=1,2,\dots,N$,
with $p_{wb}$ defined in appendix A as a function of the energy density $\epsilon$. We have called this
special initial condition isotropic water bag (iwb), (they had been independently introduced in
Ref. \cite{baldo}). For $N\rightarrow \infty$ any realization of (wb)
should not differ appreciably from (iwb). We have found that this particular realization of the water bag
conditions does not produce time fluctuations of the dynamical temperature. Moreover, in this case, the QSS temperatures do not depend on $N$ and are very close to the large $N$ value of the average QSS temperature
observed in the case of the mostly used (wb) initial conditions \cite{note}.

Besides the absence of fluctuations, it is remarkable that, at variance with the (wb), the
QSS arising from the (iwb) initial conditions behave very similarly to those produced by the two classes
of initial conditions that we have studied, in which the initial magnetization $m$ is set equal to zero.
In fact, the isotropic water bag (iwb), the uniform (un) and triangular (tr) initial conditions all share
the following properties: i) the magnetization in the QSS is given,
as a function of $N$, by $m \simeq 2N^{-1/2}$ (thus it is asymptotically zero for large $N$), and,
consequently, $T \simeq 2\epsilon-1 +4/N$; ii) the initial velocity distribution evolves in a short
time towards an attracting distribution, whose shape is approximately a semi-ellypse (see Fig. 4). 

The velocity distribution funtion becomes maxwellian when eventually the system goes to BG equilibrium,
leaving the QSS. Following this observation, one could represent the function $h(p)$, during the QSS,
exactly with a semi-ellipse (that will therefore be equal to the (el) initial conditions described in
appendix A). The two parameters (i.e. the semi axes of the ellipse) are fixed by the energy density and
by the normalization condition, and the shape is fixed. It is not surprising that, as we show in appendix B,
at $\epsilon=0.69$ the elliptic velocity distribution function satisfies the condition for its stability
as a stationary solution of the Vlasov equation \cite{yama,inag}, if $\epsilon \ge 0.625$. In
Ref. \cite{yama} a similar attracting distribution was found, although it was not parametrized as an
ellipse and its Vlasov stability through Eq. (\ref{stvl}) was not studied. It should be remarked that
the velocity distribution function evolves in time. Nevertheless, during the QSS, the evolution is very
slow, and the elliptical representation will remain a good approximation until
the relaxation toward BG equilibrium starts, when tails in the distribution begin to develop.
\vskip 0.5cm
\begin{figure}[htbp!]
\begin{center}
\includegraphics[scale=0.3]{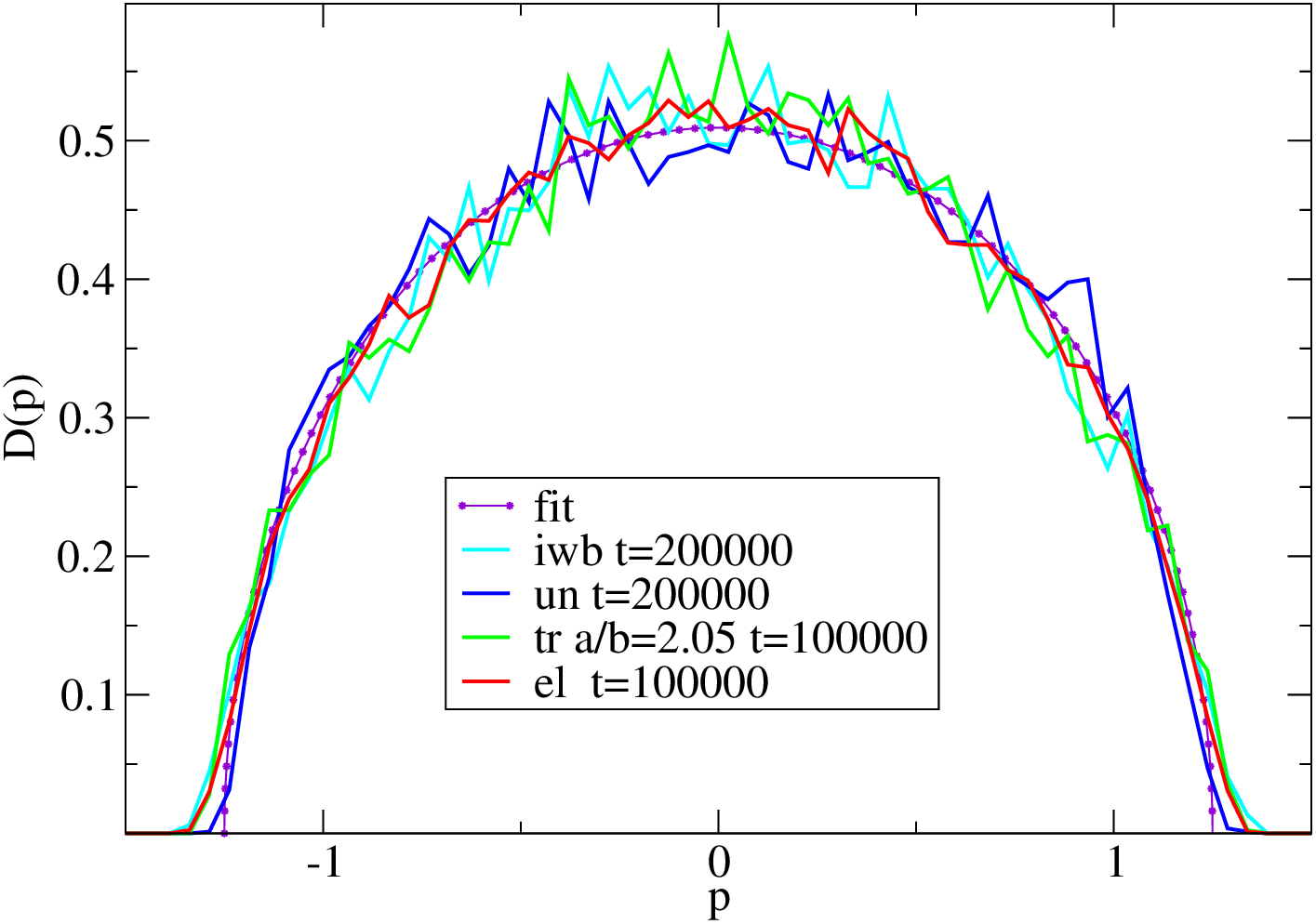}
\caption[]
{(Color online) Long time velocity distribution function for different initial conditions (see appendix A
for the meaning of the acronyms). In this case $N=16384$, $\epsilon=0.69$. For the (tr) initial conditions
we give the ratio of the distribution parameters.}
\end{center}
\end{figure}
It is then clear that, if the elliptical velocity distribution is chosen as the initial condition
(the (el) initial conditions of appendix A), the system will be prepared in a slowly
evolving QSS, without any initial transient characterized by a shorter time scale.

In the following Section we show, in particular, the occurrence of the QSS with elliptical velocity
distribution function when the dynamics starts from the (iwb) initial conditions. We then study the lifetime
of this QSS as a function of $N$, introducing a new determination of this dependence on the basis of the
fluctuations of the phase of the magnetization. Furthermore, we will show that QSS are present also
at supercritical energy densities.

\section{The dynamics of the modulus and of the phase of the magnetization}
\subsection{Relaxation dynamics of the magnetization and parallel evolution of the velocity distribution}

In this subsection we follow the time evolution of the modulus of the magnetization, in parallel with
the evolution of the one-particle distribution function $f(\theta,p)$ and of its integral over the positions,
i.e., the velocity distribution function. This is done for the (iwb) initial conditions at $\epsilon=0.69$,
with $N=10000$. As we pointed out, in this case the equilibrium magnetization is equal to $0.31$.

We show in Fig. 5 the dynamics up to time $10^6$. The quasi-stationary state is characterized by a small
value of the magnetization around $0.02$, that persists until times of about $4\cdot 10^5$,
when the relaxation toward equilibrium starts. As shown in the inset of Fig. 5 the magnetization
modulus, initially equal to $1$, falls to small values in O(1) time, with some bounces before setting to
the above mentioned value of about $0.02$. The arrows in Fig. 5 denote the times at which the snapshots
shown in Fig. 6 are taken. This figure shows the evolution of the one-particle distribution function
$f(\theta,p)$ and of the velocity distribution function. The first one is represented plotting on the
$(\theta,p)$ plane the canonical coordinates of all the rotators. One can see that during most of the
duration of the QSS, namely from a time of about $10^4$ up the time when relaxation to BG equilibrium
begins, around $4\cdot 10^5$, the velocity distribution function is characterized by the elliptical shape,
illustrated in the previous Fig. 4. The QSS ends when the semi-elliptical distribution starts to develop
tails and eventually becomes a maxwellian. Correspondingly, the developing finite magnetization
can be spotted in the dishomogeneous appearance of the left plot of panel h) of Fig. 6.
\begin{figure}[htbp!]
\begin{center}
\includegraphics[scale=0.3]{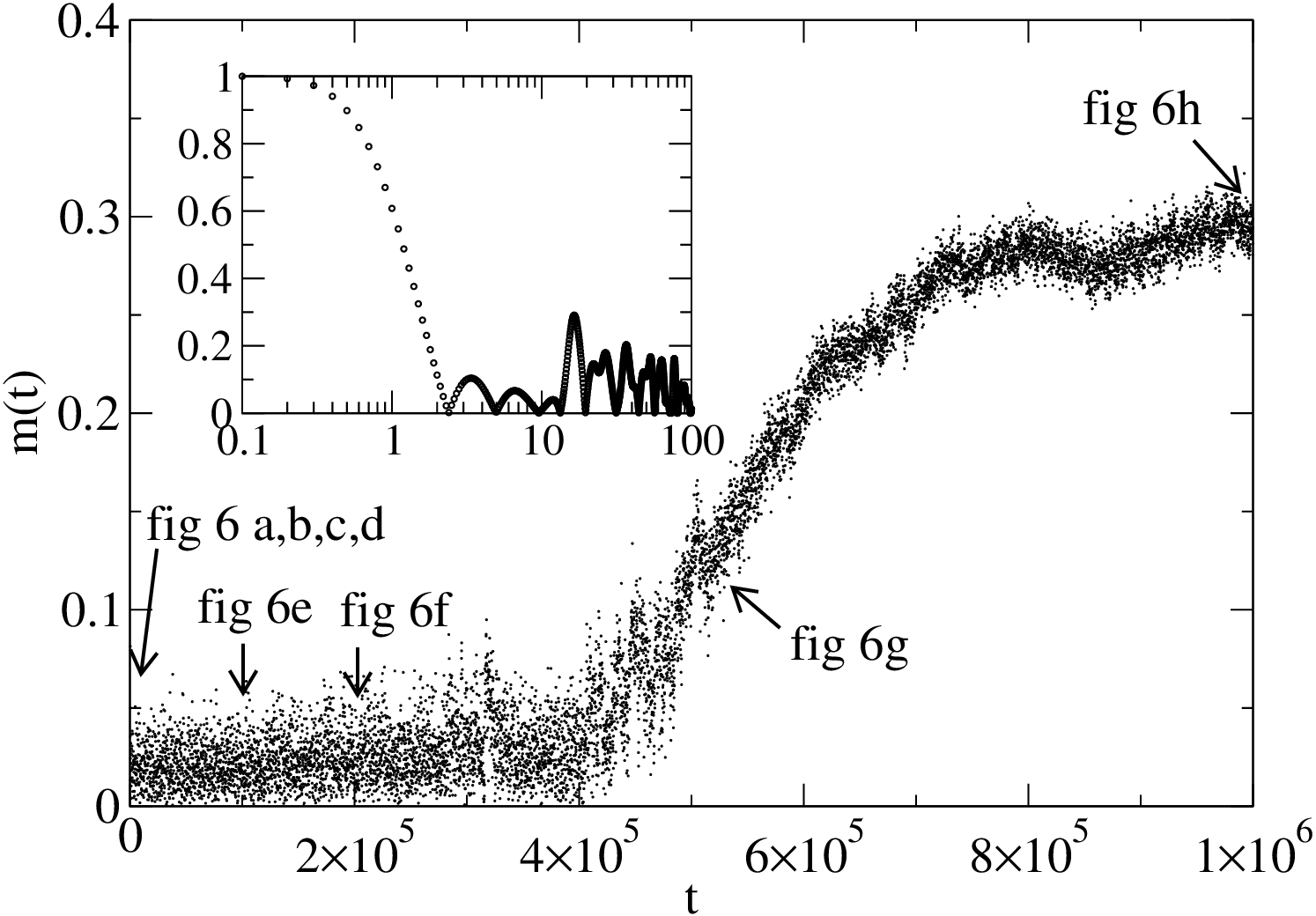}
\caption[]
{Time evolution of the modulus of magnetization of the HMF model, (iwb) initial conditions.
Note that for this class of initial conditions the initial magnetization abruptly falls on a very small
value plateau, which tends to zero, increasing $N$, approximately as $N^{\frac{1}{2}}$. A null magnetization
is associated to the quasi-stationary state. In the inset: details of the short time behaviour. Data refer
to the case : $N=10000$, $dt=0.1$, $\epsilon=0.69$. For this number of rotators the lifetime of the
quasi-stationary state is around $4\cdot 10^5$, then there is a transient ending around $7\cdot10^5$, then
the equilibrium value for the actual value of the energy density is reached: $|\mathbf{M}|=0.31$. The
arrows indicate times at which shapshots of the $\mu$-space are shown in the corresponding panels of the
following Fig. 6.}
\end{center}
\end{figure}

When the dynamics starts with the (un), (tr), and (el) initial conditions, the evolution of the
velocity distribution functions is practically the same as that presented in Fig. 6 (in the last
(el) case there is not even an initial transient). In these cases the magnetization $m$ is
practically $0$ from the beginning.
\begin{figure}[htbp!]
\begin{center}
\includegraphics[scale=0.6]{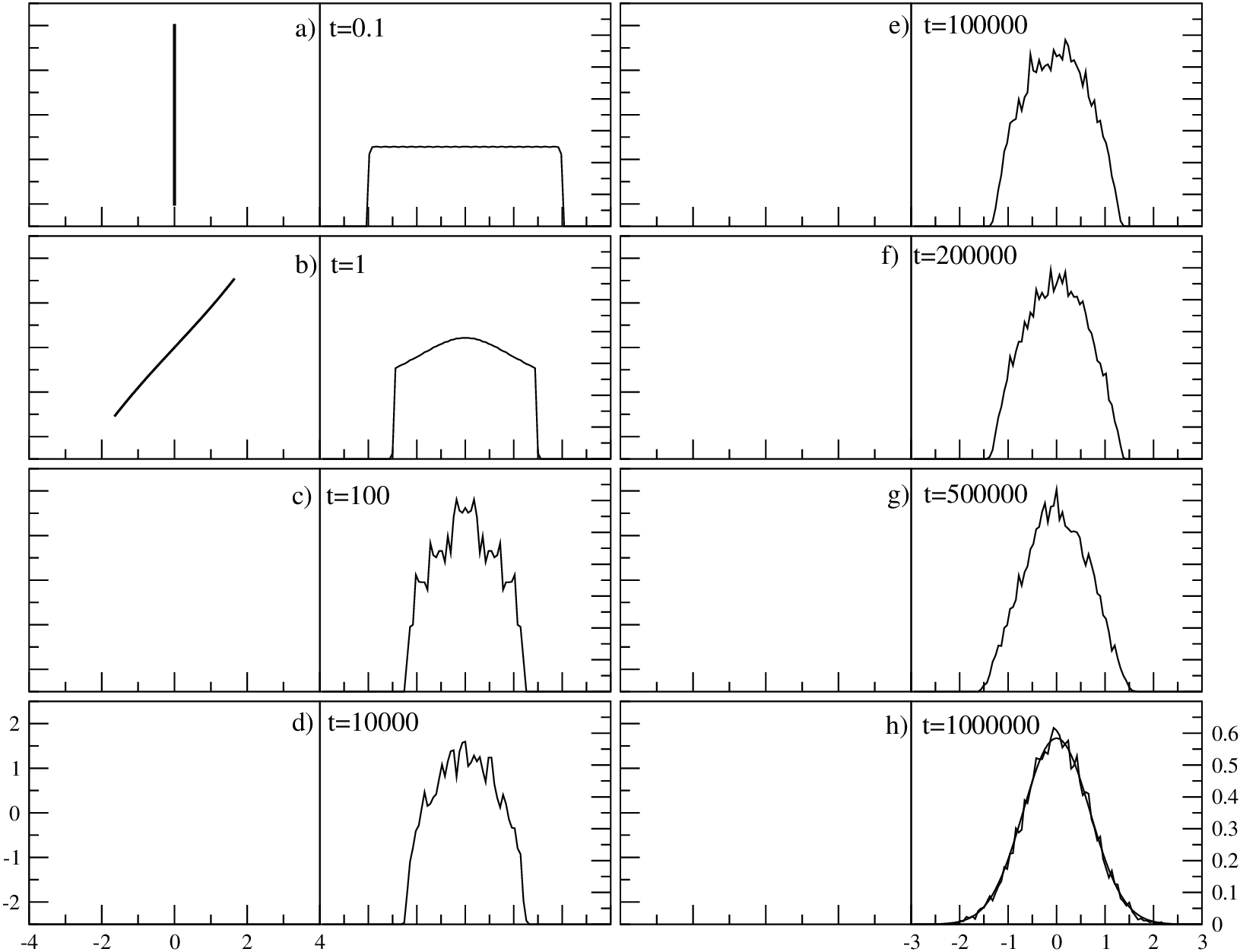}
\caption[]
{Time evolution of the $\mu-$space and, in parallel, of the distribution of velocities, (iwb) initial
conditions. The run is the same reported in Fig. 5: $N=10000$, $dt=0.1$, $\epsilon=0.69$, where the
corresponding times are marked by arrows. Clearly, after a very short initial time the velocity distribution
reaches a shape that is easily fitted to an elliptic profile. The elliptic distribution is maintained
until the dynamics is in the quasi-stationary state. After, approximately, $t=4\cdot 10^5$ the distribution
starts to develop tails and, after a transient, it becomes evidently Gaussian, in the equilibrium state.
Data in panel h) have been fitted to the maxwellian: $y=C\cdot exp(-\frac{x^2}{2T})$, the fitted temperature
results $0.475$, as expected.}
\end{center}
\end{figure}

The previous results indicate that the QSS arising from these four classes of initial conditions can be
described by as an almost-zero magnetization state, characterized by a semi-elliptical velocity
distribution function. In the next subsection the QSS is further characterized by a new quantity, the
angular frequency of the magnetization, determined by the dynamics of the argument of the magnetization. 

\subsection{Quasi-stationary states and angular frequency of the magnetization}

In the QSS the argument $\phi$ of the (very small in modulus) magnetization displays a strongly fluctuating
behaviour, corresponding to frequent and abrupt changes in direction. During the relaxation towards
equilibrium, while the modulus increases towards its equilibrium value, the fluctuations are much less
violent. This different behavior is plausible, considering the large ratio between the modulus of the
equilibrium magnetization and that of the QSS magnetization. We found that this difference can be exploited
to give a convenient characterization of the QSS. In particular, indicating with $dt$ the integration
timestep, we define the following cumulative quantity:
\begin{equation}
\sigma_m(t) = \sum_{n=0}^{t/dt} \left|\arg\left[\mathbf{m}\left(\left(n+1\right)dt\right)\right]
-\arg\left[\mathbf{m}\left(n dt\right)\right] \right|.
\end{equation}
This is the sum of the absolute values of the angular distances spanned by the magnetization vector during
an integration time $dt$, sampled up to time $t$. Let us note that this observable is monotonically
increasing with $t$, but it is also timestep dependent: for given$t$, it is monotonically not decreasing if
the sampling interval $dt$ is decreased. Fig. 7, that shows
both the modulus $m(t)$ and the the quantity $\sigma_m(t)$, refers to a typical trajectory started from the
(el) initial conditions, i.e. directly from the QSS. From the plot it is evident that there is a crossover
between two regimes, passing from the QSS to the BG equilibrium state. The crossover is characterized by the
change in the derivative $\omega_m(t)$ of $\sigma_m(t)$. We find that it is possible to fit the observed
time evolution of $\sigma_m(t)$ with:
\begin{figure}[htbp!]
\begin{center}
\includegraphics[scale=0.3]{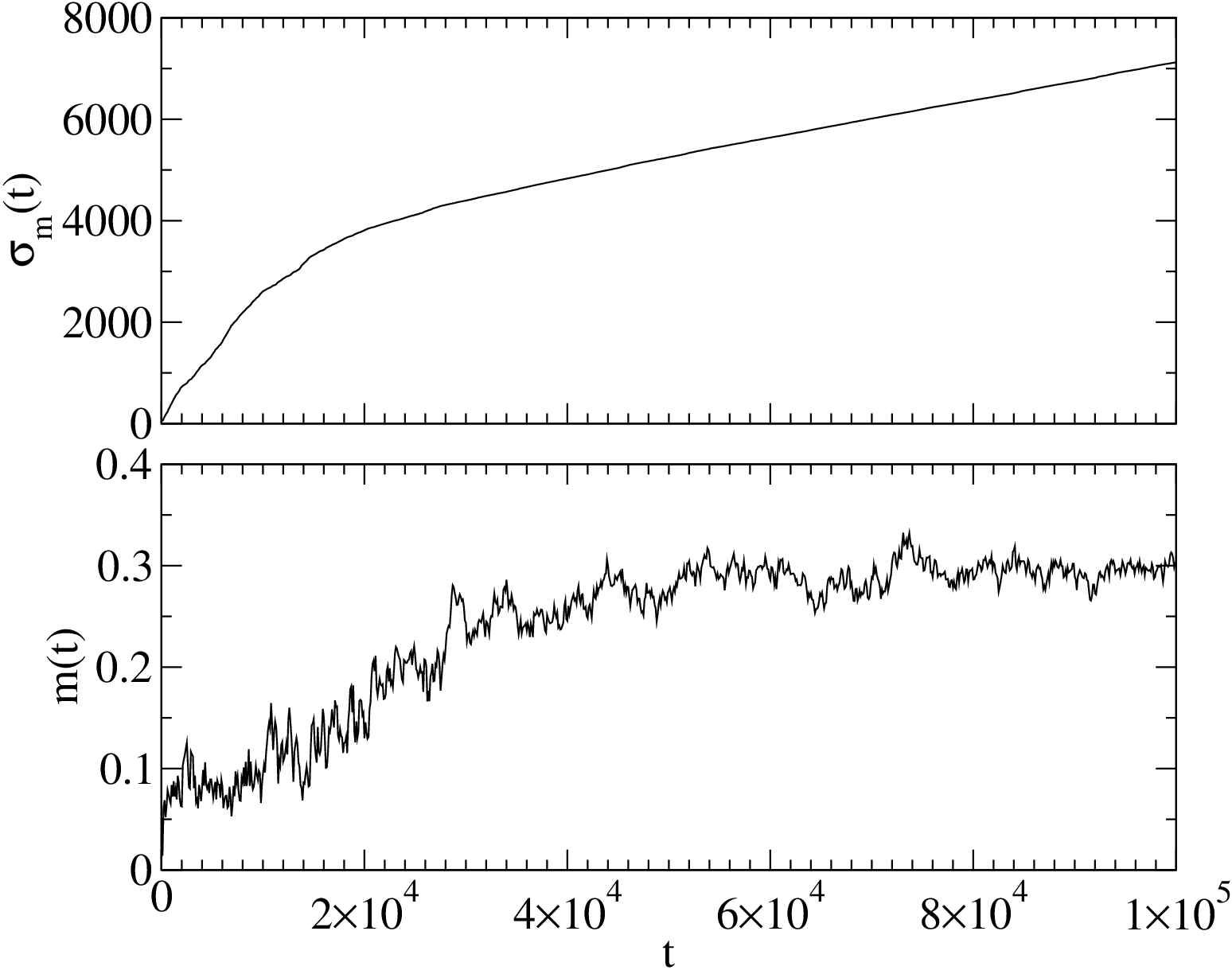}
\caption[]
{Time evolution of the function $\sigma_{m}(t)$ and of the modulus of the magnetization, (iwb) initial
conditions. Data refer to the case $N=1024$, $\epsilon=0.69$.}
\end{center}
\end{figure}
\begin{equation}\label{sigqs}
\sigma_m(t) = \omega_{m,qs} t - C t^2  
\end{equation}
in the QSS, and with:
\begin{equation}\label{sigeq}
\sigma_m(t) = \omega_{m,eq} t 
\end{equation}
in the equilibrium state. Therefore:
\begin{equation}\label{omeqs}
\omega_m(t) = \omega_{m,qs} - 2 C t  
\end{equation}
in the QSS, and:
\begin{equation}\label{omeeq}
\omega_m(t) = \omega_{m,eq} 
\end{equation}
in the equilibrium state.
We found that $\omega_{m,qs}$, the derivative in zero of $\sigma_m(t)$, depends on the energy density, but
appears to be independent of the system size $N$; on the
contrary $\omega_{m,eq}$ decreases as $N$ increases, going as the inverse square root of $N$, while $C$
depends both on $N$ and the energy density.
Namely, it increases with $\epsilon$ and decreases with $N$. In Fig. 8 we show both $m(t)$ and $\omega_m(t)$
for a given $N$ value at $\epsilon=0.69$, while in Fig. 9 we plot, for the same energy density, the
behavior of $\omega_m(t)$ for different values of $N$. In all cases we start from (el) initial conditions.
According to Eqs. (\ref{omeqs}) and (\ref{omeeq}), and as Figs. 8 and 9 show, $\omega_m(t)$ linearly decreases
with time in the QSS, and then tends to level off, when the system reaches the equilibrium state. From
Fig. 9 it is also possible to see the mentioned independence on $N$ of $\omega_{m,qs}$, while the different
slopes prove the marked dependency on $N$ of $C$ in Eq. (\ref{omeqs}).
\vskip 0.5cm
\begin{figure}[htbp!]
\begin{center}
\includegraphics[scale=0.3]{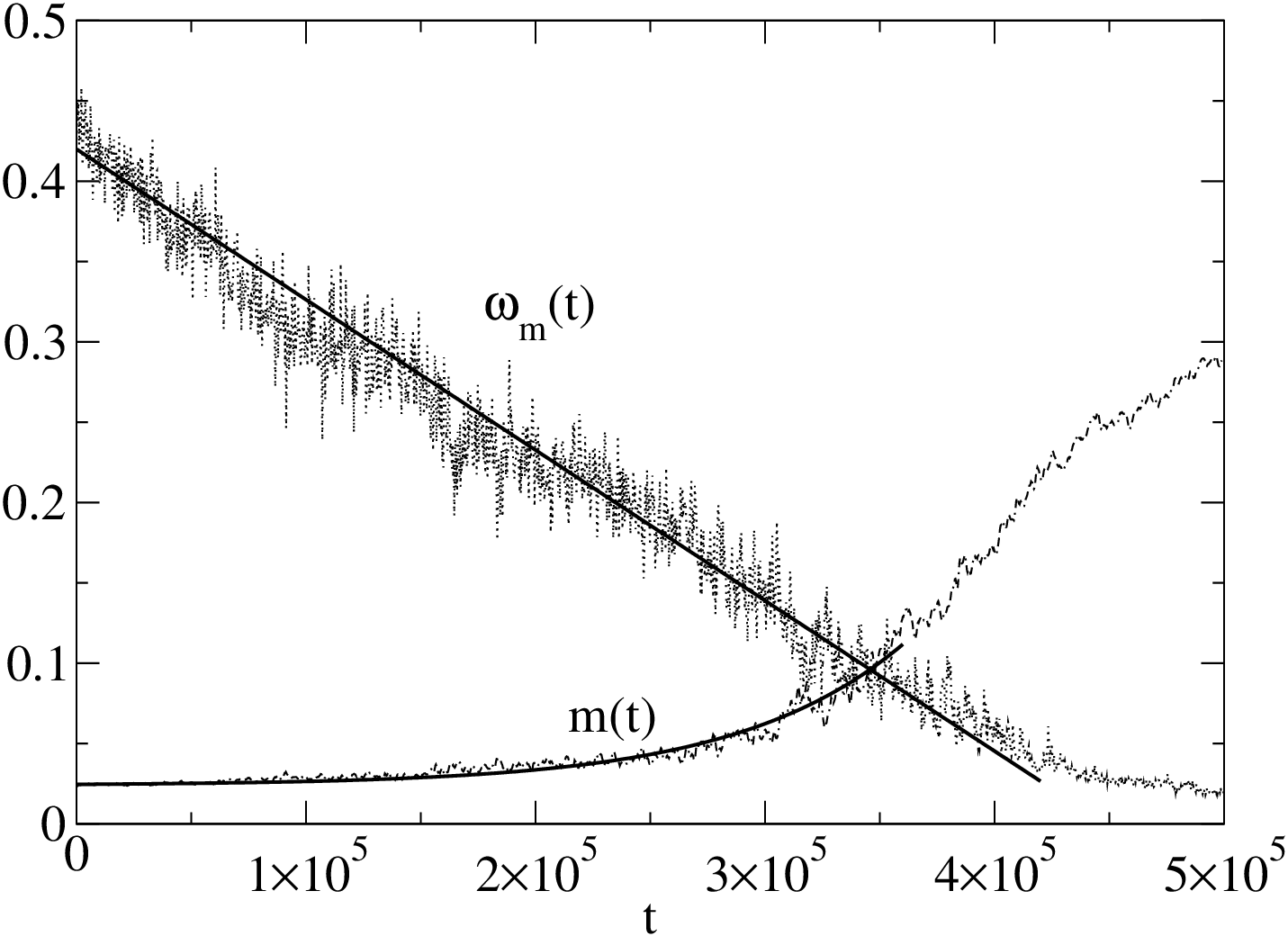}
\caption[]
{Time evolution of magnetization and of the angular frequency in a typical case: $N=8192$, $\epsilon=0.69$,
(el) initial conditions, average over 10 trajectories.}
\end{center}
\end{figure}

\begin{figure}[htbp!]
\begin{center}
\includegraphics[scale=0.3]{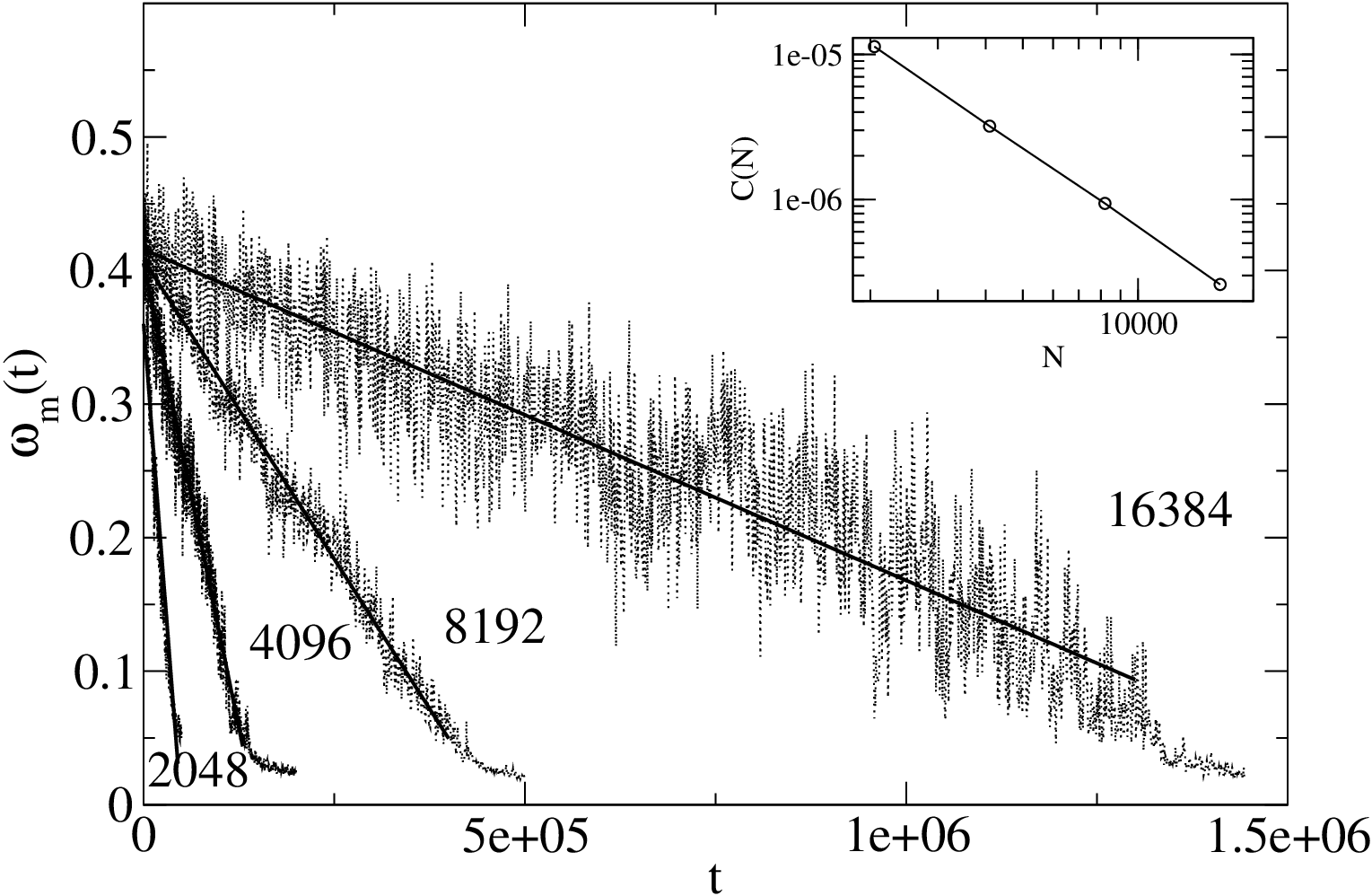}
\caption[]
{Time evolution of angular frequency of magnetization: $N$ Dependence, $\epsilon=0.69$, (el) initial
conditions. In the inset: power law dependence on $N$ of the QSS lifetime, estimated from the slope $C(N)$
of the linear regression. The fitted exponent is 1.81.}
\end{center}
\end{figure}

The change of behavior implicit in the passage from Eq. (\ref{omeqs})
to Eq. (\ref{omeeq}) seems to be much more clear cut than the gradual increase of the magnetization (or of
the temperature) at the start of the relaxation to equilibrium. Therefore, the time evolution of $\omega_m(t)$
can be used to define the lifetime of the QSS in an easier way. In fact, let us define this lifetime by the
time in which the linear fit of $\omega_m(t)$ in the QSS extrapolates to zero. Because of the independence on
$N$ of $\omega_{m,qs}$, this intercept in inversely proportional to $C$. The inset of Fig. 9 plots the
dependence of $C$ on $N$, and from this plot it is possible to reconstruct the scaling law of the QSS
lifetime. The scaling exponent is evaluated as $1.81$, in agreement with previous determinations.

The angular frequency of the magnetization, introduced here, provides a very convenient operational definition
of the QSS. The latter have been searched for, until now, exclusively at subcritical energy densities, i.e.,
for $\epsilon \leq 0.75$. However, the observations presented in the next section, based on the
quantities just introduced, indicate that QSS can be present also at supercritical energy densities.

\section{Supercritical quasi-stationary states of the HMF}

The behavior of $\omega_m(t)$ shows that, by starting from the (el) initial conditions, the system
can be set in a QSS even at super critical energy densities. In Fig. 10 we show
the time evolution of $\omega_m(t)$ and that of $m(t)$ at the supercritical energy density
$\epsilon=$0.8. Note that, also in this case, it is possible to distinguish two regimes. The crossover is
around the time $2\cdot10^{5}$. Now $\omega_m(t)$ crossovers from a logarithmic decay with time to a
constant, when the system relaxes to BG equilibrium. In this supercritical case the equilibrium value of
the magnetization is zero and the relaxation to equilibrium is characterized by the velocity distribution
function that becomes maxwellian, thing that happens at the crossover of $\omega_m(t)$, as we have checked.
Although at the crossover the magnetization modulus should remain very small, in Fig. 10 there is trace of
a small relaxation, possible signature of a finite size effect.
\begin{figure}[htbp!]
\begin{center}
\includegraphics[scale=0.3]{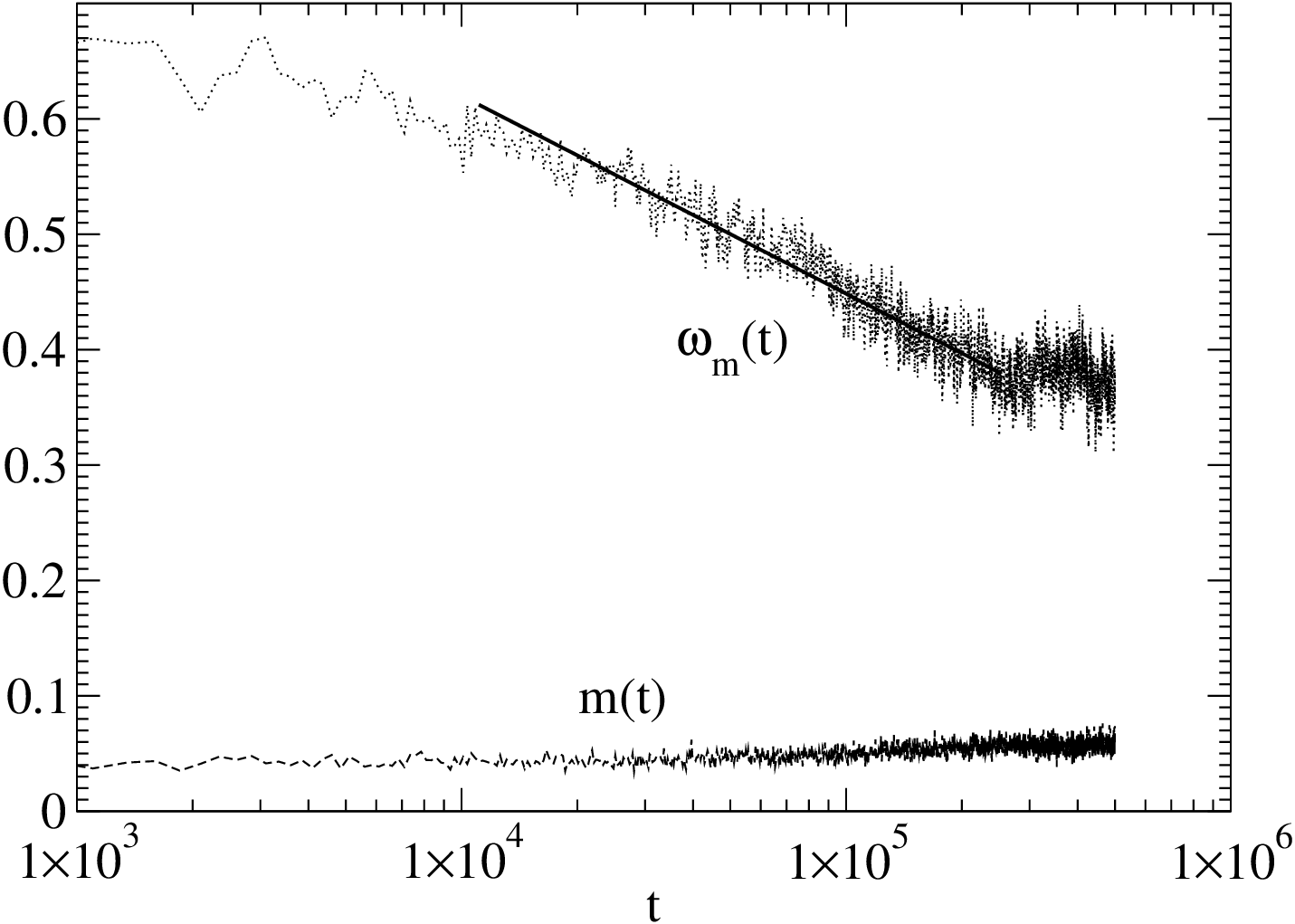}
\caption[]
{Time evolution of the magnetization and of the angular frequency of magnetization in the supercritical
case: $N=1024$, $\epsilon=0.8$, (el) initial conditions, average over 20 trajectories.}
\end{center}
\end{figure}

In Fig. 11 we show that the slope of $\omega_m(t)$ in the QSS is substantially independent of system size,
but the function is multiplied by an $N$-dependent factor, as indicated by the parallel translation of the
signal. This is the opposite behavior with respect to the subcritical case. In addition,
note the logarithmic time axis, that points, in the supercritical case, to an exponential increase of the
QSS lifetime with the system size.
\vskip 0.5cm
\begin{figure}[htbp!]
\begin{center}
\includegraphics[scale=0.3]{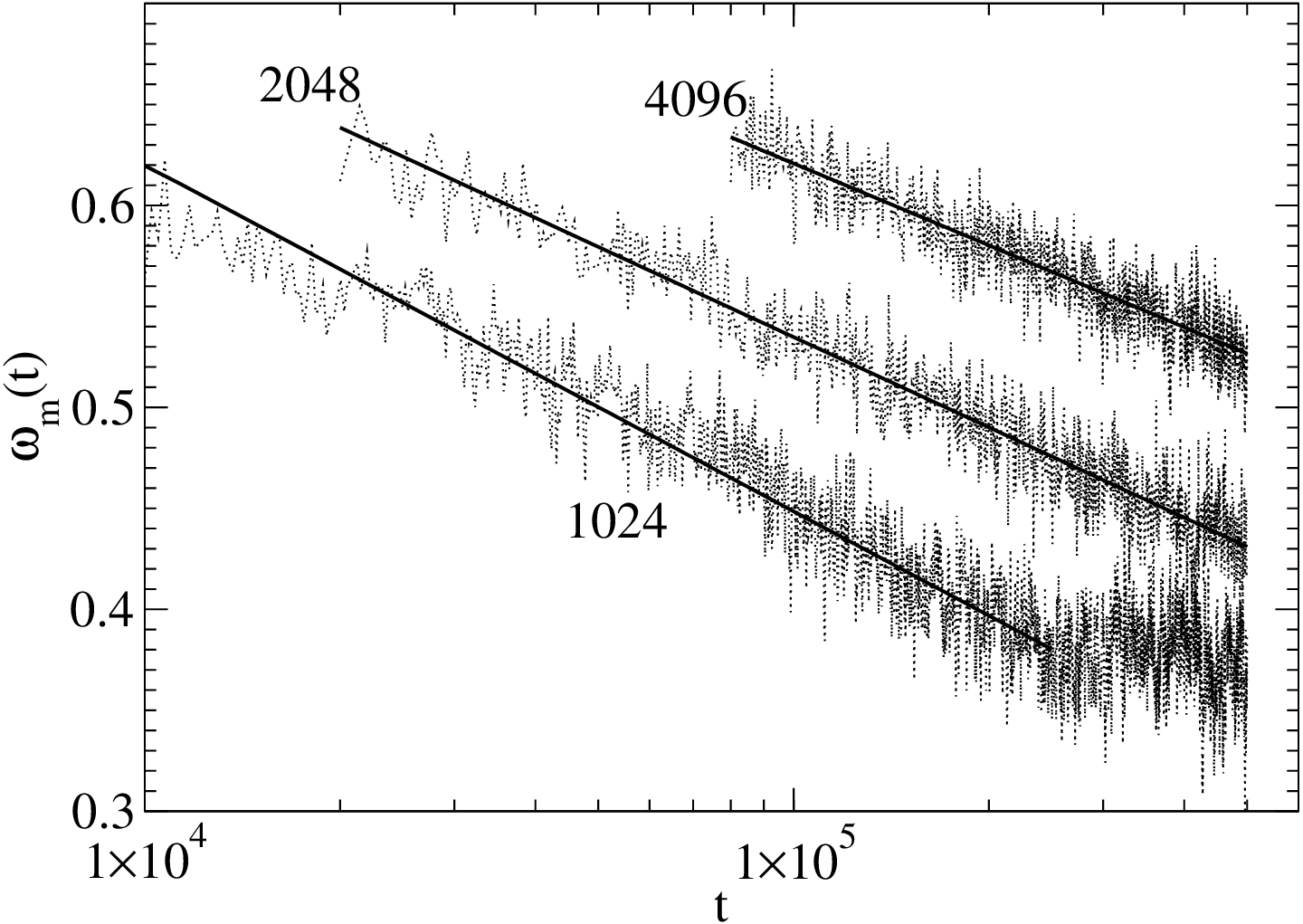}
\caption[]
{Supercritical case. $N$ dependence of the time evolution of the angular magnetization.
$\epsilon=0.8$; in each case average over 20 trajectories.}
\end{center}
\end{figure}

\section{Discussion}

In this paper we have considered different classes of initial conditions. We have observed that
at long times, before the relaxation to equilibrium, for the classes of initial conditions with $m(0)=0$ and
for the (iwb) initial conditions, which have $m(0)=1$, the velocity distribution function acquires an
elliptical shape; this is shown in Fig. 4. 
This special functional form is a stable solution of the Vlasov equation, as shown in the appendix B and in
Fig. 13. These results  are consistent with the attracting character of the elliptic velocity
distribution function.

The quasi-stationary states of the HMF have been phenomenologically characterized through the dynamics of both
the magnetization modulus and phase. These states are characterized by the small value,
$O(N^{-\frac{1}{2}})$, of the magnetization modulus and by a linear decay in time of the angular frequency of
the magnetization, an observable used here for the first time, that has been very useful to revisit the power
law behavior of the QSS lifetimes as a function of $N$. Moreover, this observable was effective in showing
that long-lasting dynamical states, quite similar to the subcritical QSS, are present also at supercritical
energy densities; it has been shown that in this case the time scale for the approach to BG equilibrium
is much greater.

Recent work on the short time behavior of the HMF model \cite{anton} has studied the dynamics on the basis
of an entropy functional which is suitable for systems in the collisionless approximation \cite{lynd}, which
is generally valid for mean field systems, and where the dynamics satisfies the Vlasov equation. Actually,
the study was restricted to the cases where the initial one-particle distribution function has only two
values: a constant in a given region of the one-particle phase space, and zero outside (i.e., the authors
consider only uniform initial distribution functions, realizing different initial magnetizations). The
dynamical mixing in the one-particle phase space will lead, in a short time scale (fast relaxation), to
a QSS, characterized by a one-particle distribution function that is a stable stationary solution of the
Vlasov equation, and that maximizes the entropy functional; the particular shape of this distribution
function depends on the initial magnetization \cite{note2}. Numerical simulations, limited to very short
times, were performed for comparison with the analytical calculation: these show that, at the
energy density $\epsilon=0.69$, the QSS has a zero magnetization, unless the initial value $m(0)$ is above
a critical value around $0.9$ \cite{anton}. Finite size effects, acting as a perturbation term in the Vlasov equation, will then be responsible, at large times, of the approach to the BG equilibrium.
Our uniform initial distribution function, with $m(0)=0$, belongs to the class studied in Ref. \cite{anton}.
In this particular case the initial distribution is already the stable stationary solution maximizing the
entropy functional, and there should be no fast relaxation. This seems to contradict our numerical results,
obtained also in Ref. \cite{yama}, showing that, when we start from the (un) initial conditions, the
distribution function has a semielliptical shape while the system is in the QSS. Then, to have another
comparison, we have investigated also a case with initial magnetization between zero and one. The
corresponding uniform initial distribution is the one called (pm) in appendix A. We have considered the
case $m(0)=0.3$, and Fig. 12 shows the distribution at different times. In this case we find that, during the
QSS, the distribution function maintains the shape reached after the first fast relaxation, that in turns
agrees with that obtained in Ref. \cite{anton}. The different behavior between the (un) and the (pm) initial
conditions can therefore be summarized in the following. The dynamics starting from the (pm) conditions
has a fast relaxation towards a QSS, with the velocity distribution slowly evolving, afterwards, to the BG
equilibrium form. Starting from the (un) conditions, where fast relaxation does not take place, the velocity
distribution has nevertheless an evolution towards the elliptical form, although quite slower than the
the fast relaxation; from there, the distribution slowly evolves towards BG equilibrium.
\vskip 0.5cm
\begin{figure}[htbp!]
\begin{center}
\includegraphics[scale=0.3]{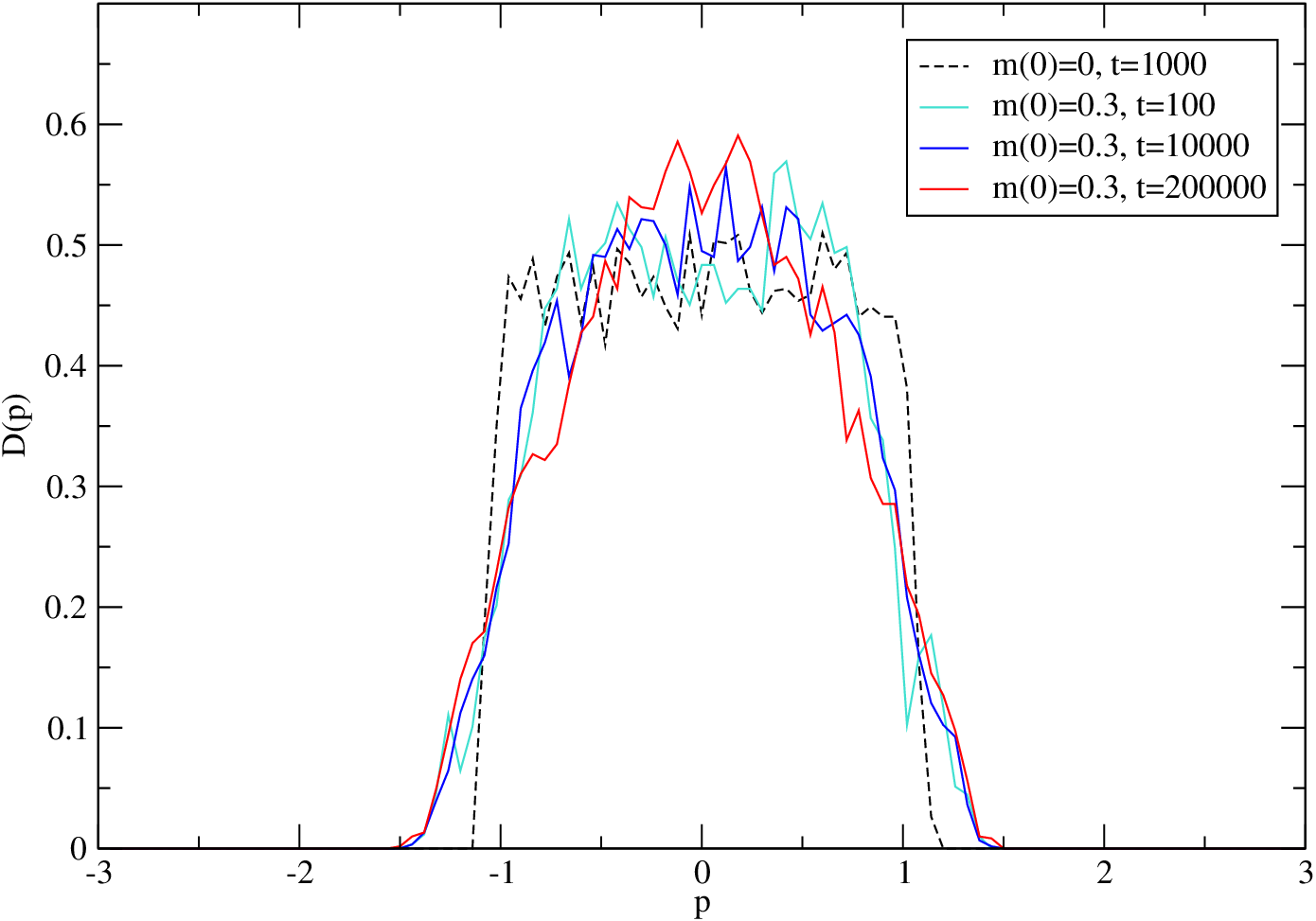}
\caption[]
{(Color online) Velocity distribution function at long times. The initial magnetization has been set
to $m(0)=0.3$; $N=10000$, $\epsilon=0.69$.}
\end{center}
\end{figure}

One can argue that a possible explanation of these different behaviors can be ascribed to a different
influence of finite size effects on both cases, according to the following.

As noted before, once in the QSS, the velocity distribution evolves very slowly, passing through a series
of stable stationary states of the Vlasov equation \cite{yama}. Finite size effects are responsible
for this very slow dynamics. As we have checked numerically, when this slow evolution drives the
velocity distribution to a situation where the stability condition (appendix B) is no more satisfied,
a faster approach to BG equilibrium takes place. However, finite size effects are present also during the
initial dynamics. They have been studied in Ref. \cite{boudau}, again on the basis of the Vlasov
equation, but without any reference to entropy functional; i.e., through a purely dynamical approach.
It has been shown that the one-particle distribution function is modified by a diffusion process.
The corresponding diffusion coefficient is proportional to $1/N$, but the proportionality coefficient
can be extremely large for the (un) initial distribution function. This would tend to modify the
distribution function rapidly. It would be interesting to perform the same calculation in the case
of the elliptic function, to have a confirmation that in this case the proportionality coefficient
is much smaller.

\begin{acknowledgments}
The present work has been supported by the PRIN05 grant on {\it Dynamics and thermodynamics of systems with
long-range interactions}. The authors thank S. Ruffo and A. Rapisarda for fruitful discussions.
\end{acknowledgments}

\appendix
\section{The distribution functions related to the different classes of initial conditions}
We collect in this appendix the distributions describing the initial conditions studied in this
paper. Let us denote with $\chi(x)$ the characteristic function of the segment $(-x,x)$, i.e., the function
which is equal to $1$ inside this segment and $0$ otherwise. The distributions $f(\theta,p)=g(\theta)h(p)$
are the following. Except in the first case, where the initial magnetization $m$ is equal to $1$, and the
third case, where it has a generic value, in the others it is equal to $0$.
\vskip 0.3cm
\noindent
$\bullet$ water bag (wb)
\begin{equation}\label{wbic}
g(\theta) = \delta (\theta) \quad \quad h(p) = \frac{1}{2p_{wb}}\chi(p_{wb})
\end{equation}
with $p_{wb}$ determined by the energy density $\epsilon$ by $p_{wb}=\sqrt{6\epsilon}$.
\vskip 0.3cm
\noindent
$\bullet$ uniform (un)
\begin{equation}\label{unic}
g(\theta)=\frac{1}{2\pi} \quad \quad h(p) = \frac{1}{2p_{un}}\chi(p_{un})
\end{equation}
with $p_{un}$ given by $p_{un}=\sqrt{6\left(\epsilon-\frac{1}{2}\right)}$; this case is possible
if $\epsilon \ge 1/2$.
\vskip 0.3cm
\noindent
$\bullet$ partial magnetized (pm)
\begin{equation}\label{pmic}
g(\theta)=\frac{1}{2\pi}\chi(\theta_m) \quad \quad h(p) = \frac{1}{2p_{pm}}\chi(p_{pm})
\end{equation}
with $p_{pm}$ given by $p_{pm}=\sqrt{6\left(\epsilon-\frac{1}{2}+\frac{1}{2}m^2\right)}$, and
$\theta_m$ by the solution of the equation $\left(\sin(\theta_m)/\theta_m\right)=m$; this case is
possible, for a given value of $m$, if $\epsilon \ge 1/2 - m^2/2$.
\vskip 0.3cm
\noindent
$\bullet$ triangular (tr)
\begin{equation}\label{tric}
g(\theta)=\frac{1}{2\pi} \quad \quad h(p) = \left(b-\frac{b-a}{p_{tr}}|p|\right)
\chi(p_{tr})
\end{equation}
where the parameters $p_{tr}$, $a$ and $b\ge a$ satisfy the two relations $(b+a)p_{tr}=1$
and $(b+3a)p_{tr}^3=12(\epsilon - \frac{1}{2})$; this case is possible if $\epsilon \ge 1/2$.
Differently from the other cases considered in this work, at a given energy $\epsilon$ there remains
a free parameter. The form of the distribution function $h(p)$ is that of a box
surmounted by a triangle.
\vskip 0.3cm
\noindent
$\bullet$ semi-elliptical (el)
\begin{equation}\label{elic}
g(\theta)=\frac{1}{2\pi} \quad \quad h(p) = \frac{2}{\pi p_{el}}\sqrt{1-\frac{p^2}{p_{el}^2}}
\chi(p_{el})
\end{equation}
with $p_{el}$ given by $p_{el}=\sqrt{8\left(\epsilon-\frac{1}{2}\right)}$; this case is possible
again if $\epsilon \ge 1/2$.

We have also considered (wb) initial conditions where the corresponding function $h(p)$ is not
realized through the usual random number generations, but the $N$ initial velocities are given the values
$p_i = -p_{wb}+2p_{wb}\left(i-1/2\right)/N$, $i=1,2,\dots,N$. This special initial condition has been called
isotropic water bag (iwb). Losely speaking, for $N\rightarrow \infty$ any realization of (wb) should tend
to (iwb). 

\section{The stability of the elliptic velocity distribution function as a stationary solution of the
Vlasov equation}
The Vlasov equation for the one-particle distribution function $f(\theta,p,t)$ of the HMF system
is given by \cite{yama,inag}:
\begin{equation}\label{vleq}
\frac{\partial f}{\partial t} + p\frac{\partial f}{\partial \theta} - \frac{\partial U}{\partial \theta}
\frac{\partial f}{\partial p}=0 \, ,
\end{equation}
where $U$ is actually a function of $(\theta,t)$ and a functional of $f$ given by:
\begin{equation}\label{ufun}
U = - \int {\rm d}\alpha{\rm d}p \cos(\theta - \alpha) f(\alpha,p,t) \, .
\end{equation} 
\vskip 0.5cm
\begin{figure}[htbp!]
\begin{center}
\includegraphics[scale=0.3]{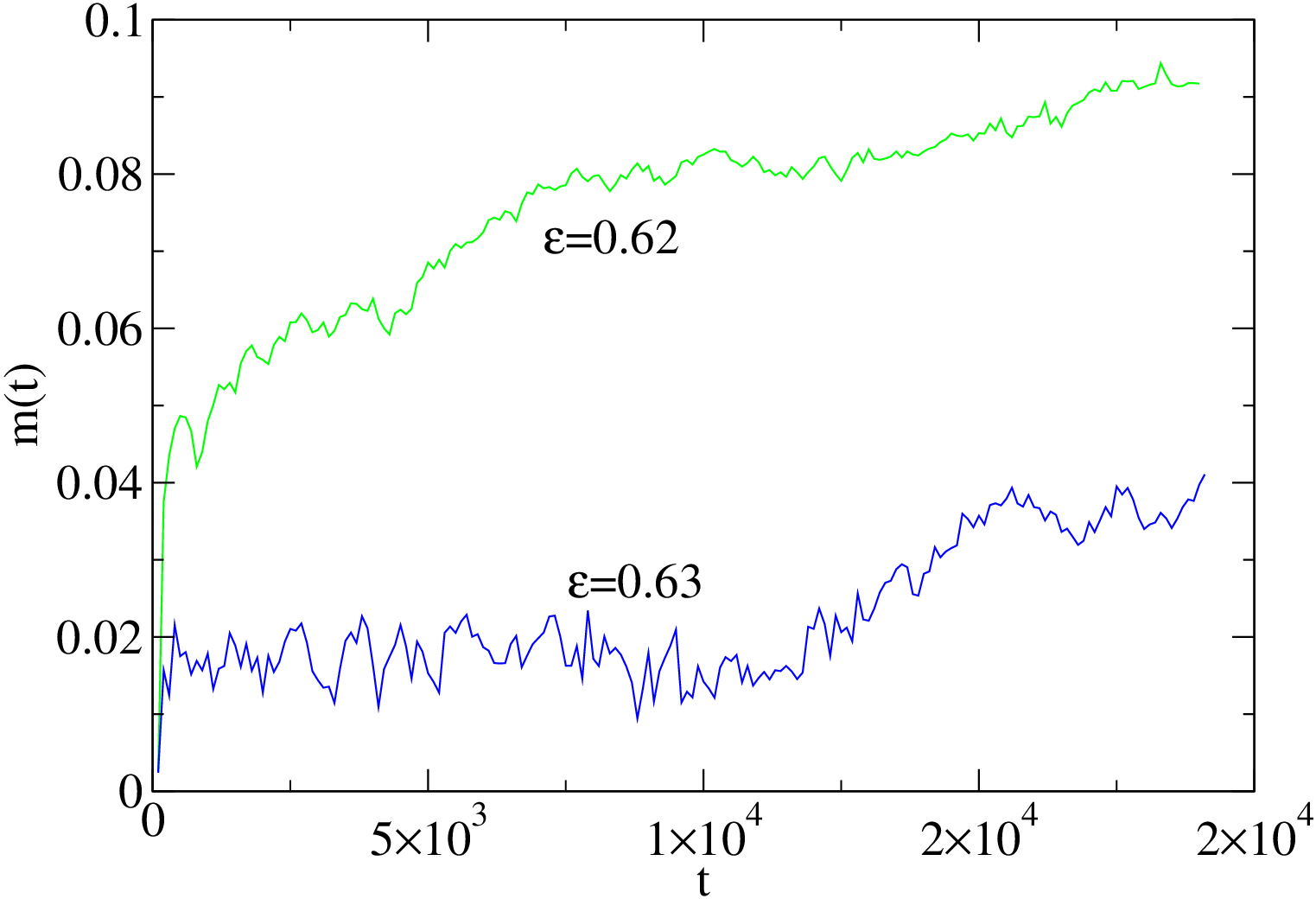}
\caption[]
{(Color online) Stability of the elliptic distribution. These simulations are for $N=262144$,
(el) initial conditions and two energy densities. Note that $\epsilon=0.62$ is below the
stability threshold $0.625$ and there is no stationary state.}
\end{center}
\end{figure}
It is immediate to see that any distribution function which is homogeneous in $\theta$, i.e., any
distribution of the form $f(\theta,p)=h(p)/(2\pi)$, is a stationary solution of the Vlasov
equation. The necessary condition for its stability can be expressed using the normalized distribution
$h(p)$; the condition is \cite{yama,inag}:
\begin{equation}\label{stvl}
1+\frac{1}{2}\int_{-\infty}^{+\infty}\frac{h'(p)}{p}{\rm d}p\geq 0 \, .
\end{equation}
In the case of the (el) initial conditions, from the expression for $h(p)$ given in appendix A we obtain:
\begin{equation}\label{stvlel}
1-\frac{1}{\pi p_{el}^{3}}\int_{-p_{el}}^{p_{el}} \frac{{\rm d}p}{\sqrt{1-(\frac{p}{p_{el}})^2}}\geq 0\, ,
\end{equation} 
that, after integration, gives: $1-\frac{1}{p_0^2}\geq 0$. Using the relation between $p_{el}$ and $\epsilon$
given in appendix A, we obtain that stability requires $\epsilon \geq \frac{5}{8}=0.625$. Therefore
the elliptic velocity distribution function is unstable if the energy density is below 0.625. Fig. 13
numerically confirms, in a concrete case, this fact.


\begin{thebibliography}{99}
\bibitem{note}
It should be noted that the symmetry of the equations of motion would imply that with the (iwb)
initial conditions the $y$ component of the magnetization $m$ should rigorously remain fixed at $0$;
however, roundoff errors in the numerical simulations do not allow this conservation.
\bibitem{note2}
Actually, the entropy functional is a conserved quantity of the Vlasov equation, and therefore the
Vlasov dynamics can not change its value from the initial one; in particular, the functional
will not be maximized by the dynamics. However, due to the mixing property of the dynamics, the functional
will be maximized by any coarse grained distribution, i.e., by the physically observable distribution.
\end{thebibliography}
\end{document}